\documentclass[sn-apa]{sn-jnl}

\usepackage{graphicx}%
\usepackage{multirow}%
\usepackage{amsmath,amssymb,amsfonts}%
\usepackage{amsthm}%
\usepackage{mathrsfs}%
\usepackage[title]{appendix}%
\usepackage{xcolor}%
\usepackage{textcomp}%
\usepackage{manyfoot}%
\usepackage{booktabs}%
\usepackage{algorithm}%
\usepackage{algorithmicx}%
\usepackage{algpseudocode}%
\usepackage{listings}%

\usepackage{natbib}

\begin{document}

\title[Article Title]{What Explains Teachers' Trust of AI in Education across Six Countries?}


\author*[1]{\fnm{Olga} \sur{Viberg}}\email{oviberg@kth.se}
\equalcont{These authors contributed equally to this work.}
\author[2]{\fnm{Mutlu} \sur{Cukurova}}\email{m.cukurova@ucl.ac.uk}
\author[3]{\fnm{Yael} \sur{Feldman-Maggor}}\email{yael.feldman-maggor@weizmann.ac.il}
\author[3]{\fnm{Giora} \sur{Alexandron}}\email{giora.alexandron@weizmann.ac.il}
\author[4]{\fnm{Shizuka} \sur{Shirai}}\email{shizuka.shirai.cmc@osaka-u.ac.jp}
\author[4]{\fnm{Susumu} \sur{Kanemune}}\email{kanemune@gmail.com}
\author[5]{\fnm{Barbara} \sur{Wasson}}\email{barbara.wasson@uib.no}
\author[5]{\fnm{Cathrine} \sur{Tømte}}\email{cathrine.tomte@uia.no}
\author[6]{\fnm{Daniel} \sur{Spikol}}\email{ds@di.ku.dk}
\author[7]{\fnm{Marcelo} \sur{Milrad}}\email{marcelo.milrad@lnu.se}
\author[2]{\fnm{Raquel} \sur{Coelho}}\email{r.coelho@ucl.ac.uk}
\author*[8]{\fnm{René F.} \sur{Kizilcec}}\email{kizilcec@cornell.com}
\equalcont{These authors contributed equally to this work.}

\affil*[1]{\orgname{KTH Royal Institute of Technology}, \orgaddress{\street{Lindstedsvagen 3}, \city{Stockholm}, \postcode{10044},  \country{Sweden}}}

\affil[2]{\orgname{University College London}, \orgaddress{\street{20 Bedford Way}, \city{London}, \postcode{WC1H 0AL}, \country{UK}}}

\affil[3]{\orgname{Weizmann Institute of Science}, \orgaddress{\street{Herzl St 234}, \city{Rehovot}, \postcode{7610001}, \country{Israel}}}

\affil[4]{\orgname{Osaka University}, \orgaddress{\street{1-1 Yamadaoka}, \city{Osaka}, \postcode{565-0871}, \state{Suita}, \country{Japan}}}

\affil[5]{\orgname{University of Bergen, SLATE}, \orgaddress{\street{Christies gate 12}, \city{Bergen}, \postcode{5007}, \country{Norway}}}

\affil[6]{\orgname{University of Copenhagen}, \orgaddress{\street{Nørregade 10}, \city{København}, \postcode{1172}, \country{Denmark}}}

\affil[7]{\orgname{Linnaeus University}, \orgaddress{\street{Universitetsplatsen 1}, \city{Växjö}, \postcode{35252}, \country{Sweden}}}

\affil[8]{\orgname{Cornell University}, \orgaddress{\street{107 Hoy Rd}, \city{Ithaca}, \postcode{14853}, \state{NY}, \country{USA}}}


\abstract{With growing expectations to use AI-based educational technology (AI-EdTech) to improve students' learning outcomes and enrich teaching practice, teachers play a central role in the adoption of AI-EdTech in classrooms. Teachers' willingness to accept vulnerability by integrating technology into their everyday teaching practice, that is, their trust in AI-EdTech, will depend on how much they expect it to benefit them versus how many concerns it raises for them. In this study, we surveyed 508 K-12 teachers across six countries on four continents to understand which teacher characteristics shape teachers' trust in AI-EdTech, and its proposed antecedents, perceived benefits and concerns about AI-EdTech. We examined a comprehensive set of characteristics including demographic and professional characteristics (age, gender, subject, years of experience, etc.), cultural values (Hofstede's cultural dimensions), geographic locations (Brazil, Israel, Japan, Norway, Sweden, USA), and psychological factors (self-efficacy and understanding). Using multiple regression analysis, we found that teachers with higher AI-EdTech self-efficacy and AI understanding perceive more benefits, fewer concerns, and report more trust in AI-EdTech. We also found geographic and cultural differences in teachers' trust in AI-EdTech, but no demographic differences emerged based on their age, gender, or level of education. The findings provide a comprehensive, international account of factors associated with teachers' trust in AI-EdTech. Efforts to raise teachers' understanding of, and trust in AI-EdTech, while considering their cultural values are encouraged to support its adoption in K-12 education.}


\maketitle

\section{Introduction}

Over the last decade, artificial intelligence-based educational technology (AI-EdTech) has been increasingly used in K-12 (kindergarten, primary and secondary) education across many countries~\citep{UNESCO2023gem,holmes2022artificial,zawacki2019systematic}. Its use has accelerated since 2019 across grade levels~\citep{crompton2022affordances} and evolved towards the end of 2022 with the rapid integration of generative AI technologies due to the sudden popularity of OpenAI's ChatGPT~\citep{lim2023generative}. AI-EdTech can potentially improve students' learning outcomes and processes, critical thinking, and problem-solving skills~\citep{wu2022effectiveness}, for example, by providing more immediate feedback and personalized learning paths~\citep{benotti2017tool}. AI-EdTech also promises benefits for teachers. According to a recent systematic review, these include improved planning (e.g., around students' needs), implementation (e.g., immediate feedback and teacher intervention), and assessment (e.g., through automated essay scoring)~\citep{celik2022promises}. Such benefits have the potential to improve the conditions for student learning and ultimately lead to improved learning outcomes. However, our understanding of the impact of AI-EdTech on teaching and learning is in its infancy as there is not yet widespread adoption of AI-EdTech in schools and a lack of empirical studies of its use~\citep{holmes2022artificial}.

A critical area for systematic investigation is teachers' perspectives on AI-EdTech, as they are ultimately the ones who need to meaningfully integrate it into their routines in ways that improve student learning and support their everyday teaching practices~\citep{kizilcec2023advance, seufert2021technology}. The decision to adopt a new technology (whether it be a new curriculum or an AI-EdTech) creates vulnerability for the teacher because it might not work as expected and cause problems that the teacher will have to resolve, which may have negative consequences for the teacher's performance and relationship with the students. The willingness to take on this vulnerability is fundamentally an act of trust in the AI-EdTech~\citep{hosmer1995trust}. Although teachers are essential stakeholders for AI-EdTech, our scientific understanding of their trust in this technology, and which individual and contextual factors influence their trust, is limited~\cite[e.g.,][]{velander2023artificial,cukurova2023adoption}. Cukurova and colleagues~\citeyearpar{cukurova2023adoption} note that the "slow adoption of AIED systems in real-world settings might, in part, be attributable to the frequent neglect of a range of other factors associated with complex education systems" (p. 152).

Prior research has examined several factors influencing teachers’ adoption of AI-EdTech in education. These include teachers’ level of knowledge about AI~\citep[e.g.,][]{velander2023artificial}, perceived self-efficacy, anxiety, perceived usefulness, perceived ease of use~\citep[e.g.,][]{almaiah2022measuring, wang2021factors}, and trust in these technologies~\citep[e.g.,][]{nazaretsky2022teachers, nazaretsky2022instrument}. Prior work highlights that a teacher's decision to adopt a new technology, such as AI-EdTech, will be influenced by their trust in the technology~\citep{moore1991development}. In particular, as Agarwal and Prasad~\citeyearpar{agarwal1997role} point out, individuals "will be less likely to experiment with new technologies if they perceive a significant risk associated with such exploration" (p.574). In other words, they are unwilling to make themselves vulnerable because of their perceived concerns with the technology---an important antecedent of trust. What factors influence teachers' trust in AI-EdTech, including their proposed antecedents, perceived benefits and concerns of AI-Edtech, has received limited research attention, especially at scale and across countries. Only recently have researchers established frameworks for assessing teachers' perceived benefits, concerns, and trust related to AI-EdTech~\citep{choi2023influence, nazaretsky2022teachers, nazaretsky2022instrument, al2023acceptance}. This has created a foundation for studying factors that influence teachers' adoption of AI-EdTech in schools around the world.

Multi-national studies are necessary in this domain because individuals' attitudes towards technology adoption are known to differ considerably across countries~\citep{huang2019cultural, morrone2009good}, and people's trust in automation can be influenced by cultural differences~\citep{berkovsky2018cross, chien2016relation, huang2018users}. In their seminal review of 82 articles on culture in information systems research, Leidner and Kayworth~\citeyearpar{leidner2006review} found that several dimensions of culture influence the adoption and use of information technology. Their analysis is grounded in Hofstede’s model of cultural values~\citep{hofstede2010cultures}, a widely used model to conceptualize and study culture. Their findings show that four cultural values, namely \textit{uncertainty avoidance}, \textit{power distance, masculinity }(vs. femininity), and \textit{collectivism }(vs. individualism), are important predictors of technology adoption (see Background for details). Context (e.g., cultural values, educational level, etc.) is known to play an important role in assessments of trust~\citep{hong2019cross, klein2019trust}, and to our knowledge, there is no prior research on the impact of culture on school teachers' trust in AI-EdTech.

To improve our understanding of factors that influence teachers’ adoption of AI-EdTech, the present study investigates teachers’ trust in AI-EdTech, and two of its antecedents, perceived benefits and concerns about AI-EdTech, in the context of K-12 education in six countries (Brazil, Israel, Japan, Norway, Sweden, USA). Based on prior work by Hofstede and colleagues~\citeyearpar{hofstede2010cultures}, we selected countries to represent distinct geographic areas (North Europe, East Asia, Middle East, North and South America), in which individuals' cultural values are expected to differ along several dimensions (e.g., uncertainty avoidance and power distance). We also note that according to the Oxford Insights Government AI Readiness Index~\citep{aiready2022}, these countries differ in terms of their overall readiness to utilize AI as of 2022. Considering the importance of cross-cultural differences in people's attitudes about automation and ICT (information and communication technologies), this study aims to understand better how teachers' trust in AI-EdTech varies across countries and cultural value dimensions. This study marks an important step towards understanding the teacher perspective that can inform the development of interventions to support teachers’ effective adoption and use of AI in schools to realize its full potential. Our study addresses the following research question:\\
\textbf{RQ:} To what extent is K-12 teachers' trust in AI-EdTech explained by (a) their perceived benefits and concerns about AI-EdTech, (b) demographic and professional characteristics, (c) AI self-efficacy and understanding, (d) cultural values, and (e) geographic location?

\section{Background}

\subsection{Teachers' Trust in AI-EdTech}

We adopt the classic definition of trust as the willingness to accept vulnerability to another entity~\cite{hosmer1995trust}. It is a necessary condition for any type of cooperative behavior~\citep{davis1962externalities}, including the formation of interpersonal relationships~\citep{slovic1993perceived} and the adoption of a technology in teaching~\citep{nazaretsky2022teachers}. Trust is recognized as ``a critical aspect of AI adoption and usage''~\citep[p. 1994]{lukyanenko2022trust}. With the rise of AI, the issue of trust in AI has become especially important, and despite scholars’ increased attention to the topic, evidence in this area remains fragmented, especially in K-12 education settings~\citep{kizilcec2023advance}. The 2022 \textit{European Commission Ethical Guidelines on the use of AI and data in teaching and learning} explicitly identified the promotion of excellence and trust in AI as a priority for the Commission~\citep{eceg2022}. Few evidence-based research studies have examined teachers’ trust in AI-EdTech thus far. Lukyanenko and colleagues~\citeyearpar{lukyanenko2022trust} note, ``trust in the broader systems [e.g., K-12 education], in which AI is embedded due to AI being a component of the system, has, so far, escaped much research'' (p.2008). This critical observation motivates the present study. 

In their seminal model of trust, Mayer and colleagues~\citep{mayer1995integrative} identify three causal antecedents of trust: competence, benevolence, and integrity. Applied to the context of technology, these causal antecedents have been interpreted as beliefs about (1) the functionality of the technology, (2) its helpfulness, and (3) its reliability~\citep{mcknight2011trust}. In the context of teachers' trust in AI-EdTech, these antecedents need to be examined more succinctly and naturally as beliefs about benefits (functionality and their helpfulness) and concerns (reliability). In fact, a recently proposed survey instrument for measuring trust in AI-EdTech developed subscales specifically about the perceived benefits of AI-EdTech and concerns about using AI-EdTech~\citep{nazaretsky2022instrument}. Perceived benefits and concerns about AI-EdTech have also been highlighted as important antecedents in prior work. For example, a study of teachers’ perceptions of AI as a tool to support teaching in Estonian K-12 classrooms highlights that teachers recognize possible benefits of AI technology for learning and productivity, but raise concerns about its reliability and how it could undermine interpersonal communication and obstruct social components of learning~\citep{chounta2022exploring}. Concerns about AI have also been framed more broadly as \textit{algorithm aversion} in AI-adoption research in other domains~\citep{dietvorst2015algorithm}.

\begin{figure}[htbp]
  \centering
  \includegraphics[width=\textwidth]{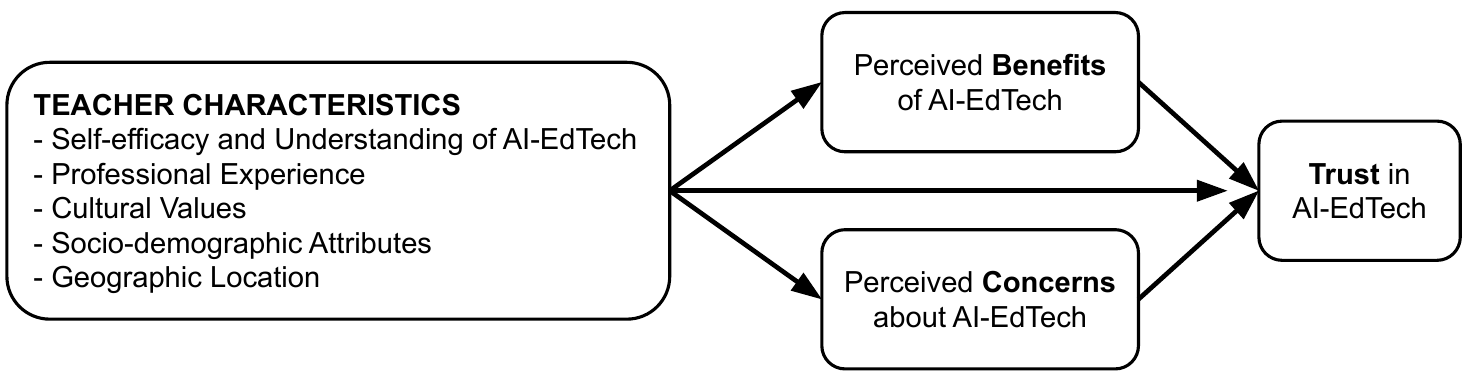}
  \caption{Overview of the relationships that are investigated in this study.}
  \label{fig:schema}
\end{figure}

We consider these two constructs---perceived benefits and concerns---as key antecedents of trust in this research, as shown in the diagram in Figure~\ref{fig:schema}. In this study, we investigate the relationship between trust and its antecedents, and teacher characteristics that explain variation in each of these constructs. We examine a wide variety of teacher characteristics, many of which have been highlighted in prior work as sources of variation in teachers' trust. 

\subsection{Sources of variation in teachers' trust in AI-EdTech}

Prior work has examined several factors as potential explanations for variation in teachers' attitudes toward adopting AI-EdTech in K-12 education. Although these studies are conducted in different geo-cultural contexts and with different methodological approaches, they observe similar patterns of results. They highlight the importance of perceived benefits and limited concern to develop trust in AI-EdTech platforms and ultimately encourage their adoption. In a survey of 792 grade 5-12 teachers in the United Arab Emirates, the factors that correlated with real-world use of an adaptive learning platform included teachers' trust and perceived ownership of the adaptive platform, workload considerations, perceived ease of use, and community support~\citep{cukurova2023adoption}. Likewise, a survey study of 83 science teachers in Abu Dhabi found that AI-EdTech acceptance is associated with teachers' perceived benefits, ease of use, and self-efficacy beliefs~\citep{al2023acceptance}. Another survey of 140 Estonian K-12 teachers found that their knowledge about AI and how it could support them is limited, and they require more support to understand how to use it effectively~\citep{chounta2022exploring}. In addition to teachers' perceived technology competence, a study of 187 mathematics teachers in Malaysia found that teachers' willingness to use AI-EdTech was also influenced by the school's culture regarding technology and access to support and resources~\citep{ayub2015factors}.

Across several studies, \textit{self-efficacy} beliefs are hypothesized as a key determinant of trust among teachers, and self-efficacy has been found to influence trust in AI in a teacher-focused study~\citep{nazaretsky2022teachers}. \cite{bandura1977self} defined self-efficacy as confidence in one's ability to perform a task or undertake an endeavor. Self-efficacy has been found to influence persistence tasks and interest expression~\citep{evans2014role}. The development of self-efficacy requires experiences overcoming obstacles through continuous effort~\citep{usher2008sources}. Prior experience can therefore help build self-efficacy and foster trust: for example, in a study about trust and experience with driving automation, experienced drivers tended to be less sensitive to risk and more trusting of automation~\citep{he2022modelling}. This presents an intervention opportunity to enhance prior experience and understanding directly. Nazaretsky and colleagues~\citeyearpar{nazaretsky2022teachers} developed and evaluated a teacher professional development program (PDP) to increase K-12 teachers’ trust in AI-EdTech. The PDP helped teachers gain important knowledge about AI-powered assessment, which helped forestall some of their misconceptions and biases related to the nature of AI-EdTech. By the end of the program, teachers expressed a higher willingness to use AI-EdTech in their classrooms (i.e., more trust in it) and proposed innovative ways to integrate this tool into their pedagogy. They were also more open to data-driven decision making in education; this can reinforce their positive attitudes toward AI. All this highlights the importance of improving an understanding of AI among teachers to develop trust and acceptance.

Teachers' attitudes towards and their self-efficacy with AI-EdTech have been examined through various theoretical models, constructs, and operationalization, including the Technology Acceptance Model (TAM)~\citep{al2023acceptance}, the TPACK framework~\citep{celik2023towards}, and several newly developed instruments~\citep{cukurova2023adoption, chou2023level}. The diversity in theoretical and methodological approaches raises issues for generalizing findings across studies. Although prior studies examined teachers' attitudes toward adopting AI-EdTech in several countries, there are few cross-cultural analyses to test if results generalize across contexts. Moreover, while several studies examined the nature of teachers' attitudes towards AI-EdTech~\citep[e.g.,][]{chounta2022exploring,celik2023towards}, fewer studies focus on exploring different factors, especially their trusting beliefs, that might influence them. This calls for more large-scale, cross-cultural research on teachers' trust and factors influencing trust. The present research addresses this call by studying teachers' \textit{trust} in AI-EdTech and its two antecedents, \textit{perceived benefits} of and \textit{concerns} about AI-EdTech, across several countries. We also study previously overlooked factors that may influence their trust in AI-EdTech, including teacher characteristics such as their years of teaching experience and years of experience using EdTech in teaching, and individual cultural values such as long-term orientation and uncertainty avoidance.




\subsection{Cultural differences in technology adoption and trust in EdTech}

Cultural values differ across countries and influence people's responses to their environment~\citep{hofstede2010cultures}, including how much they trust people and institutions~\citep[e.g.,][]{doney1998understanding, hofstede2010cultures, klein2019trust} as well as technology and automation~\citep[e.g.,][]{berkovsky2018cross, chien2015cross, huang2018users, yerdon2017investigating}. As highlighted by Chien and colleagues~\citeyearpar{chien2015cross}, "cultural values and norms can greatly influence an individual’s trust and reliance on the automation as well as the formation, dissolution and restoration of trust" (p.688).

Although research on AI-EdTech is increasingly conducted around the world, the role of culture in teachers' trust in AI-EdTech is relatively understudied in the broader literature on culture in education. Relevant research in other contexts studied different types of information and communication technologies, including AI tools. For example, \cite{chien2016relation} examined the effects of cultural characteristics (i.e., the dimensions of power distance, individualism, and uncertainty avoidance, and personality traits) on reported trust in automation in Taiwanese, Turkish, and US samples. They found that uncertainty avoidance and individualism were significantly correlated with respondents' general trust in automation in the US and Turkish sample, but not in the Taiwanese sample. In another study, \cite{berkovsky2018cross} examined cultural differences when using recommendation systems across four countries (France, Japan, Russia, and the US). Their experiments with 102 participants from these countries found cultural differences in preferences for information presentation and explanation based on nine constructs of trust (e.g., transparency, integrity, etc.). They used Hofstede's original scores for the masculinity dimension to interpret their results~\citep{hofstede2010cultures}; a culture’s masculinity score (e.g., high in Japan) was strongly correlated with participants' preference for human (vs. AI) presentation and personalized explanations because they had more trust in them. Finally, Yerdon and colleagues~\citeyearpar{yerdon2017investigating} examined cultural differences in trust in the setting of automotive automation, specifically for autonomous driving. They found that a driver's expected communication style, which varied based on the cultural dimension of individualism/collectivism, significantly affected their level of trust. Together, these studies demonstrate the importance of cultural differences in people's trust in different technologies and forms of automation.


The present study employs Hofstede’s theoretical model of national culture \citep{hofstede2010cultures} to examine the possible influence of teachers' cultural values on their perceived benefits, concerns, and trust in AI-EdTech. Hofstede and colleagues~\citeyearpar{hofstede2010cultures} define culture as ``the collective programming of the mind which distinguishes the members of one group or category of people from another'' (p.5). While our study is grounded in Hofstede’s model of culture and its core dimensions (individualism/collectivism, masculinity/femininity, power distance, uncertainty avoidance, and long/short-term orientation), we do not assume homogeneity in cultural values within a nation. Instead, we use these cultural dimensions to examine \textit{teacher's individual cultural values}, and how they relate to their perceived benefits, concerns, and trust in AI-EdTech. We chose to adopt Hofstede's model as it can ``explain human behaviors better than other measures, such as country and language'' (\citep{li2022cross}, p. 269). It has previously been used to study cultural differences in students' and teachers' attitudes towards technology use in educational settings~\citep{huang2019cultural, viberg2013cross, tarhini2017examining}, as well as trust in information systems, including AI technologies~\citep{berkovsky2018cross}, and automation across countries~\citep{chien2018effect}. 

\section{Methods}

\subsection{Context and Participants}

We collected responses from school teachers in six countries. Table~\ref{tab:desc} provides a comprehensive overview of sample characteristics. The data collection process (described below) varied across countries to adhere to local research standards. We obtained approval from institutional or national ethics boards before data collection, and all respondents included in the analysis provided informed consent. Teachers' participation was voluntary, and data collection occurred between November 2022 and June 2023.

The Israeli sample was collected from middle and high school science, math, and computer science teachers. The survey was distributed among in-service teachers’ professional development communities and teachers studying for a master’s degree in science education. Teachers completed the survey using an online platform built on WordPress. The survey items for measuring teachers' perceived benefits, concerns, and trust in AI-EdTech were originally composed in Hebrew and validated by \citep{nazaretsky2022instrument}. The culture-related items were first translated into Hebrew and validated by the Israeli research team and the five additional science education experts. The research received ethical approval from an Institutional Review Board (IRB) committee (code number 11862).

The Norwegian sample was collected at five upper secondary schools in Norway's east, south, and western parts. All selected schools are medium to large schools in terms of the number of students, spanning from about 300 to 1000. School principals were sent information about the study and they invited their teachers to participate in the survey. Principals either sent the researchers a list of the teacher's e-mail addresses and they were sent the survey link, or they distributed the link to the survey directly to their teachers. After a short introduction to the study, the researchers involved, and the right to withdraw, teachers provided consent for the data to be used for research and proceeded to the survey. The online survey was administered using the software SurveyXact, which allowed for anonymous participation. The researchers translated the survey instrument into Norwegian, and several rounds of editing were conducted. The agency reported and approved the study for privacy and ethical concerns in research at the University of Bergen.

The Swedish sample was collected from junior and senior upper secondary school teachers. The survey was distributed among in-service teachers through professional development communities and the schools' principles, as well as through a newsletter sent by the Swedish National Agency of Education in May 2023. The survey instrument was translated into Swedish and validated by the Swedish research team, the two additional experts and the two teachers who piloted the survey items. The survey was distributed through an online platform, Artologik Survey and Report. Ethical approval was obtained from the National Ethical Review Board (2023-00291-01) in March 2023.

The Brazilian sample was collected from secondary school teachers using social media channels, and through the mailing list of a research center focused on basic education school teachers in Brazil. In addition to sending the survey link to teachers' school email lists, the professional network of the Regional Center for Studies on the Development of the Information Society was approached. The survey was cross-translated to Portuguese, and teachers completed the survey online using the Qualtrics survey platform. The survey had a preamble section informing the teachers about the study, the purpose of the research, the team involved, and the right to withdraw at any time. Teachers who agreed to proceed were asked to consent to data collection and use for research purposes. The study protocol received ethics approval from the institutional ethics board (REC 1760: Trust in AI-based EdTech and Cultural Values) and the data protection registration (Z6364106/2023/01/153).

The US teacher sample was collected by asking a superintendent to distribute the survey among teachers in a school district located in Upstate New York. A short study description was provided along with the URL to Qualtrics survey. The description read ``Research about AI in Classrooms (short survey): The Future of Learning Lab, as part of an international research effort to support teaching practices, is inviting high school science teachers to complete a 15-20min survey about their attitudes toward AI-based technology for teaching. The study will compare teachers’ responses across [countries].'' The institutional review board at Cornell University approved the study protocol.

The Japanese sample was collected from junior high and high school teachers in science, mathematics, and computer science. Using Microsoft Forms, the anonymous online survey was distributed to several junior high and high school teachers' networks in Japan, where the co-authors are based. All the original questionnaire items in English were translated into Japanese and validated by the Japanese research teams. The research received ethical approval from an IRB committee (code number 2022-24).

\begin{sidewaystable}
\centering
\footnotesize
\caption{Descriptive statistics for each country's sample and the combined sample: percentages for categorical variables; means with standard deviations in parentheses for continuous variables.}\label{tab:desc}
\begin{tabular}{lrrrrrrr}
  \toprule
  & Brazil & Israel & Japan & Norway & Sweden & USA & Overall \\ 
  \midrule
  Sample Size & 41 & 99 & 87 & 102 & 111 & 68 & 508 \\ 
  Trust in AI-EdTech M (SD) & 3.69 (0.64) & 3.71 (0.60) & 3.35 (0.47) & 2.98 (0.69) & 3.21 (0.59) & 2.93 (0.69) & 3.29 (0.68) \\ 
  Perceived Benefits M (SD) & 4.02 (0.56) & 3.97 (0.72) & 3.61 (0.60) & 3.56 (0.82) & 3.77 (0.65) & 3.43 (0.84) & 3.71 (0.74) \\
  Perceived Concerns M (SD) & 2.91 (0.74) & 3.14 (0.68) & 3.06 (0.46) & 3.68 (0.54) & 3.33 (0.58) & 3.66 (0.58) & 3.33 (0.64) \\ 
  AI-EdTech Self-efficacy M (SD) & 2.46 (0.74) & 1.89 (0.88) & 1.73 (0.71) & 2.51 (0.94) & 2.42 (0.89) & 2.01 (1.02) & 2.17 (0.93) \\  
  AI Understanding M (SD) & 0.90 (0.85) & 0.98 (0.73) & 0.44 (0.90) & 0.91 (0.76) & 0.91 (0.75) & 0.98 (0.87) & 0.85 (0.82) \\ 
  Uncertainty Avoidance M (SD) & 3.93 (0.58) & 3.39 (0.70) & 3.30 (0.48) & 3.70 (0.75) & 3.65 (0.66) & 3.91 (0.59) & 3.61 (0.68) \\ 
  Masculinity M (SD) & 1.95 (0.97) & 1.89 (0.90) & 2.11 (0.77) & 2.05 (1.12) & 1.63 (0.72) & 1.71 (0.74) & 1.88 (0.90) \\ 
  Collectivism M (SD) & 3.19 (0.77) & 3.04 (0.75) & 2.63 (0.56) & 3.39 (0.82) & 2.87 (0.73) & 2.79 (0.69) & 2.98 (0.77) \\ 
  Long-term Orientation M (SD) & 3.97 (0.61) & 4.06 (0.51) & 3.63 (0.43) & 4.10 (0.63) & 4.07 (0.46) & 3.95 (0.51) & 3.97 (0.55) \\ 
  Power Distance M (SD) & 1.68 (0.61) & 1.73 (0.54) & 1.88 (0.61) & 1.60 (0.75) & 1.54 (0.52) & 1.59 (0.53) & 1.66 (0.61) \\ 
  Age: \% 20-30 & 10 & 4 & 9 & 9 & 6 & 4 & 7 \\ 
  Age: \% 31-40 & 20 & 18 & 34 & 22 & 23 & 29 & 24 \\ 
  Age: \% 41-50 & 22 & 36 & 32 & 36 & 31 & 35 & 33 \\ 
  Age: \% 51-60 & 44 & 30 & 22 & 27 & 29 & 26 & 29 \\ 
  Age: \% 61+   & 5 & 11 & 2 & 6 & 12 & 4 & 7 \\ 
  Gender: \% M/F/Non-binary & 44/56/0 & 31/69/0 & 84/16/0 & 60/39/1 & 59/41/0 & 44/54/1 & 55/45/0 \\  
  Highest Edu.: \% Bachelor  & 17 & 15 & 53 & 34 & 68 & 3 & 36 \\ 
  Highest Edu.: \% Master    & 68 & 66 & 41 & 63 & 22 & 90 & 55 \\ 
  Highest Edu.: \% Doctorate & 12 & 19 & 6 & 2 & 3 & 6 & 7 \\ 
  Highest Edu.: \% Other     & 2 & 0 & 0 & 1 & 7 & 1 & 2 \\ 
  Subject: \% Biology     & 5 & 18 & 26 & 7 & 16 & 24 & 17 \\ 
  Subject: \% Chemistry   & 10 & 55 & 28 & 4 & 16 & 22 & 23 \\ 
  Subject: \% Mathematics & 29 & 18 & 45 & 10 & 5 & 4 & 17 \\ 
  Subject: \% Physics     & 10 & 10 & 29 & 9 & 25 & 13 & 17 \\ 
  Subject: \% CS          & 7 & 17 & 31 & 4 & 37 & 6 & 19 \\ 
  Subject: \% Other       & 59 & 27 & 23 & 76 & 74 & 31 & 50 \\  
  Experience in Educ.: \% 0-5yrs     & 22 & 18 & 7 & 24 & 10 & 7 & 14 \\ 
  Experience in Educ.: \% 6-10yrs    & 10 & 23 & 18 & 19 & 21 & 21 & 19 \\ 
  Experience in Educ.: \% 11-20yrs   & 29 & 23 & 41 & 35 & 35 & 32 & 33 \\ 
  Experience in Educ.: \% 21+yrs     & 39 & 35 & 33 & 23 & 34 & 40 & 33 \\ 
  Teaching w/ Technology: \% 0-5yrs  & 37 & 27 & 44 & 25 & 10 & 10 & 24 \\ 
  Teaching w/ Technology: \% 6-10yrs & 37 & 41 & 24 & 19 & 31 & 22 & 29 \\ 
  Teaching w/ Technology: \% 11-20yrs& 17 & 23 & 22 & 48 & 44 & 40 & 34 \\ 
  Teaching w/ Technology: \% 21+yrs  & 10 & 8 & 10 & 9 & 15 & 28 & 13 \\ 
   \bottomrule
\end{tabular}
\end{sidewaystable}

\subsection{Measures}

An anonymous online survey was distributed to K-12 teachers in different countries. The survey instrument measured the following constructs: trust in AI-EdTech, perceived benefits and concerns about AI-EdTech, and a variety of teacher characteristics. The complete questionnaire is available on OSF: \url{https://osf.io/abe9u/?view_only=b691f9365c0b4272b9472df5e81518f0}. Table~\ref{tab:desc} shows the distribution of responses in each sample and overall.

\textbf{Trust in AI-EdTech:} Trust was measured using an index of 5-point Likert scale responses to nine statements adopted from \citep{nazaretsky2022instrument}: e.g. ``I fully trust using AI-based educational technology in my classroom.'' The response scale ranged from strongly disagree (1), disagree (2), neither agree nor disagree (3), agree (4), to strongly agree (5). Responses to the nine statements were averaged into an index with high internal reliability (Cronbach's alpha = 0.806).

\textbf{Perceived benefits of AI-EdTech:} Benefits were measured using an index of seven statements adopted from \citep{nazaretsky2022instrument}: e.g. ``Artificial Intelligence can assist teachers with in-class management activities, such as identifying students who are off-task.'' Responses to the seven statements were rated on a 5-point Likert scale and averaged with high internal reliability (alpha = 0.850).

\textbf{Perceived concerns about AI-EdTech:} Concerns were measured using an index of 5-point Likert-scale responses to nine statements adopted from \cite{nazaretsky2022instrument}: e.g. ``AI algorithms do not understand social, emotional, and motivational factors that are very important in teaching.'' Responses to the nine statements were averaged into an index with adequate internal reliability (alpha = 0.764).

\textbf{AI-EdTech Self-efficacy:} Self-efficacy regarding AI-EdTech was measured using an index of two items: ``How knowledgeable are you about AI applications’ use in education?'' (5-point scale from Not knowledgeable at all [1], Slightly knowledgeable, Moderately knowledgeable, Very knowledgeable, to Extremely knowledgeable [5]), and ``How confident are you discussing AI applications in education with others?'' (5-point scale from Not confident at all [1], Slightly confident, Moderately confident, Very confident, to Extremely confident [5]). Responses were averaged into an index with high internal reliability (alpha = 0.850).

\textbf{AI Understanding:} We assessed teachers' understanding of AI by asking them to select from a set of options the ones they consider the closest description of AI. This item was inspired by previous surveys on public perceptions of AI~\citep[e.g.,][]{cave2019scary}. As there is not one correct description of AI, we provided five options that vary in how accurately they describe AI in the context of EdTech. The two options we deemed most accurate (coded as $1$) were ``an algorithm for making decisions based on big data'' and ``a relationship between concepts, problems, and methods for solving problems.'' The two options we deemed least accurate (coded as $-1$) were ``autonomous robots'' as in terminator-like science fiction and ``automated tasks that repeat themselves'' as in traditional software. One option we deemed moderately accurate (coded as $0.5$): ``a replica of human intelligence'' as in a simulation of human intelligence by machines \citep{dong2020research}. We convert responses into a score between -2 to 2.5 by adding up the coded values.

\textbf{Cultural Values (power distance, uncertainty avoidance, collectivism, long-term orientation, and masculinity):} We use the Individual Cultural Values Scale (CVSCALE)~\citep{yoo2011measuring} to measure teachers' cultural values with 26 items. This instrument is based on Hofstede's~\citep{hofstede2010cultures} cultural categories, but instead of assessing culture at the national level, it operates at the individual level, and it is valid for both ``student and non-student samples'' (p. 205) in five countries~\citep{yoo2011measuring}. Each cultural values dimension is an index of four to six statements rated on a 5-point Likert scale: strongly disagree (1), disagree (2), neither agree nor disagree (3), agree (4), strongly agree (5). Responses were averaged into an index with adequate internal reliability (power distance: alpha = 0.755; uncertainty avoidance: alpha = 0.717; collectivism: alpha = 0.803; long-term orientation: alpha = 0.718; masculinity: alpha = 0.806).

\textbf{Socio-demographic attributes:} Teachers reported their age group (20-30, 31-40, 41-50, 51-60, 61+), gender identity (M, F, non-binary), and highest level of education (Bachelor, Master, Doctorate, Other). 

\textbf{Professional Experience:} Teachers reported their primary teaching subject(s), which were grouped into the common domains (Biology, Chemistry, Mathematics, Physics, CS, and Other). They also reported their number of years of experience in education, and their number of years of experience using EdTech tools in teaching (0-5, 6-10, 11-20, 21+ years).

\begin{sidewaystable}
\centering
\footnotesize
\caption{Pearson correlation coefficients between pairs of constructs.}\label{tab:cor}
\begin{tabular}{lccccccccc}
  \toprule
  \textbf{Construct} & \textbf{2} & \textbf{3} & \textbf{4} & \textbf{5} & \textbf{6} & \textbf{7} & \textbf{8} & \textbf{9} & \textbf{10} \\ 
  \midrule
  1. Benefits of AI-EdTech              & -0.33$^{***}$ & 0.62$^{***}$ & 0.22$^{***}$ & -0.04 & 0.08 & 0.01 & 0.13 & -0.18$^{**}$ & 0.14$^{*}$ \\ 
  2. Concerns about AI-EdTech             &  & -0.50$^{***}$ & -0.14$^{*}$ & 0.00 & 0.18$^{**}$ & 0.21$^{***}$ & 0.16$^{**}$ & 0.15$^{*}$ & -0.04 \\ 
  3. Trust in AI-EdTech                &  &  & 0.14$^{*}$ & 0.10 & 0.07 & 0.07 & 0.10 & -0.03 & 0.10 \\ 
  4. AI-EdTech Self-Efficacy         &  &  &  & -0.07 & -0.04 & 0.04 & 0.04 & -0.19$^{***}$ & 0.13 \\ 
  5. Power Distance        &  &  &  &  & 0.09 & 0.18 & -0.08 & 0.43$^{***}$ & -0.13\\ 
  6. Uncertainty Avoidance &  &  &  &  &  & 0.22$^{***}$ & 0.32$^{***}$ & 0.01 & -0.01 \\ 
  7. Collectivism          &  &  &  &  &  &  & 0.33$^{***}$ & 0.17$^{**}$ & 0.05 \\ 
  8. Long-term orientation &  &  &  &  &  &  &  & 0.03 & 0.00 \\ 
  9. Masculinity           &  &  &  &  &  &  &  & & -0.09 \\ 
  10. AI Understanding        &  &  &  &  &  &  &  & &  \\
  \bottomrule
   \multicolumn{10}{l}{\textit{Note:} Statistical significance with multiple-testing adjustment indicated by $^{***} p<0.001$, $^{**} p<0.01$, $^{*} p<0.05$.}
\end{tabular}
\end{sidewaystable}

\subsection{Analytic Approach}

We visualized variation in the key constructs, trust in AI-EdTech, perceived benefits and concerns about AI-EdTech (Figure~\ref{fig:boxplot}), and examined their correlation with AI-EdTech self-effiacy, AI understanding, and cultural values (Table~\ref{tab:cor}). Then, we conducted two sets of regression analyses to systematically answer our research question. First, we fitted a multiple linear regression model for each antecedent of trust (perceived benefits and concerns), with the same set of teacher characteristics as predictors: demographic, professional, AI-EdTech self-efficacy, AI understanding, and cultural values (Table~\ref{tab:regression1}). This pair of models provided insight into what teacher characteristics explain perceived benefits and concerns, the proposed antecedents of trust in AI-EdTech. 

The second set of regression models systematically examines the relationship between trust in AI-EdTech, its two proposed antecedents, and various teacher characteristics. Across three variations of the model, trust is explained by (1) only its two antecedents, (2) only the various teacher characteristics, and (3) antecedents and teacher characteristics combined. This systematic approach provides insight into factors that explain trust in isolation of, and above and beyond the proposed antecedents.

In the regression models, to account for the geographical clustering in our data, we use country fixed effects instead of random effects (i.e. a mixed-effects model), because it guarantees statistical consistency and the key assumption that the predictors are uncorrelated with the random intercept does not hold~\citep{bell2019fixed}. In particular, we checked the models for potential multicollinearity by computing the variance inflation factor (VIF) for each predictor variable: all VIF scores are below 3, except for the sample fixed effects. This indicates that all predictor variables are sufficiently uncorrelated (VIF scores below 10 indicate low levels of multicollinearity), and there is no need to apply variable selection (i.e., it would not meaningfully change our findings). The higher VIF scores for the sample fixed effects indicate that the predictor variables are correlated with the country fixed effect, which violates a central assumption of random effects models, and leads us to use a fixed effect model.

Tables \ref{tab:regression1} and \ref{tab:regression2} show pooled regression results (over multiple imputations) with heteroskedasticity-robust standard errors. We performed ten imputations with predictive mean matching over 50 iterations to impute a small amount of missing data using the \textit{mice} package in R~\citep{van2011mice}. The share of missing values was less than 7\% of observations. Following best practices for analyzing multiply imputed data, we fit the same multiple linear regression model on each of the ten imputed datasets and pool the results. This method accounts for the uncertainty associated with imputing missing data directly in the calculation of standard errors and \textit{p}-values~\citep{van2011mice}. The results are reproducible using the analysis script and de-identified data available on OSF: \url{https://osf.io/abe9u/?view_only=b691f9365c0b4272b9472df5e81518f0}.

\section{Results}

\begin{figure}[htbp]
  \centering
  \includegraphics[width=0.99\textwidth,trim=0 20 0 0, clip=true]{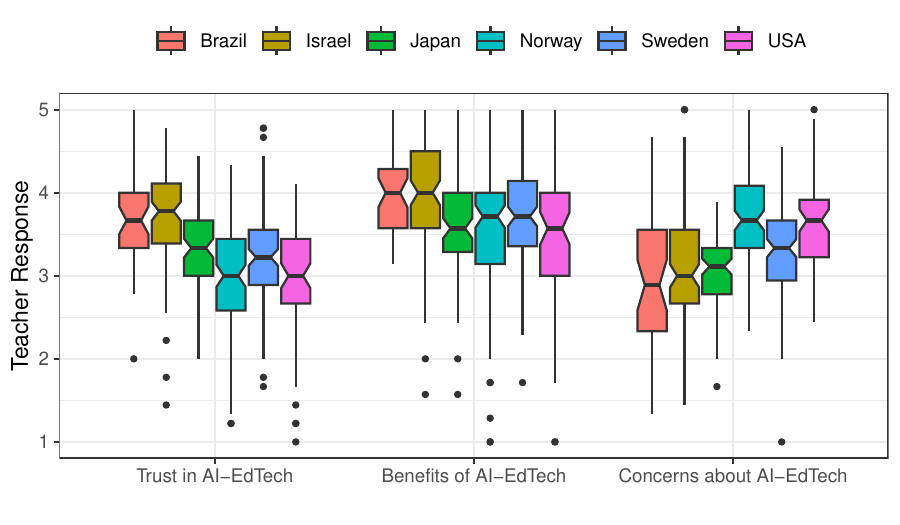}
  \caption{Variation in perceived benefits, concerns, and trust in AI-EdTech by teacher sample (n=508).}
  \label{fig:boxplot}
\end{figure}

Figure~\ref{fig:boxplot} shows the variation in teachers' perceptions of benefits, concerns, and trust in AI-EdTech within and across our six samples. Perceptions of benefits are mostly above the neutral midpoint (3 on the 1-5 scale) and follow a similar pattern as teachers' trust in AI-EdTech. Teachers' concerns are relatively high and appear inversely related to perceived benefits and trust. Formally, Table~\ref{tab:cor} indicates that the three variables---benefits, concerns, and trust---are significantly correlated with each other, and the correlations with concerns are negative. Additionally, Table~\ref{tab:cor} shows that the variables are correlated with AI-EdTech self-efficacy, AI understanding, and several cultural dimensions.

The samples collected in the six countries covered a range of demographic characteristics, academic subjects, levels of experience with teaching and technology, cultural values, and AI-EdTech self-efficacy and AI understanding (see Table~\ref{tab:desc}). We used multiple regression analysis to disentangle how these different factors are associated with teachers' trust and its two antecedents, as depicted in Figure~\ref{fig:schema}. First, we conducted two multiple regression analyses for the antecedent variables (perceived benefits and concerns of AI-EdTech) with all teacher characteristics. All predictors were entered simultaneously because there was no evidence of multicollinearity (i.e., correlations between predictors; all VIF scores $<$ 8), and thus no need to perform variable selection. The regression results, presented in Table~\ref{tab:regression1}, yield the following five main findings that are statistically significant while adjusting for all other factors.

\begin{table}[ht]
\centering
\small
\caption{Results for multiple regressions explaining perceived benefits and concerns of AI-EdTech with various teacher characteristics and fixed effects for the geographic location (sample). Regression coefficients are shown with heteroskedasticity-robust standard errors in parentheses.}\label{tab:regression1}
\begin{tabular}{lcc}
  \toprule
                            &  \textbf{Benefits of AI-EdTech} & \textbf{Concerns about AI-EdTech} \\
   \midrule
   (Intercept)              &  2.47$^{***}$ (0.43) &  2.65$^{***}$ (0.32) \\
   AI-EdTech Self-Efficacy            &  0.18$^{***}$ (0.04) & -0.13$^{***}$ (0.03) \\
   AI Understanding            &  0.11$^{*}$   (0.04) & -0.07$^{*}$   (0.03) \\
   Uncertainty Avoidance    &  0.13$^{*}$   (0.06) &  0.05     (0.04) \\
   Masculinity              & -0.10$^{*}$   (0.05) &  0.10$^{**}$  (0.03) \\
   Collectivism             & -0.03     (0.05) &  0.10$^{*}$   (0.04) \\
   Long-term Orientation    &  0.17$^{*}$   (0.07) &  0.02     (0.05) \\
   Power Distance           &  0.01     (0.07) & -0.05     (0.05) \\
   Experience in Education  &  0.01     (0.05) & -0.09$^{*}$   (0.04) \\
   Teaching with Technology &  0.03     (0.05) &  0.08     (0.04) \\
   Highest Edu.: Master     & -0.13     (0.08) &  0.03     (0.06) \\
   Highest Edu.: Doctorate  & -0.04     (0.15) & -0.08     (0.12) \\
   Highest Edu.: Other      & -0.22     (0.23) &  0.06     (0.18) \\
   Subject: Biology         &  0.03     (0.10) & -0.15     (0.08) \\
   Subject: Chemistry       &  0.05     (0.10) & -0.01     (0.08) \\
   Subject: CS              & -0.12     (0.09) &  0.02     (0.08) \\
   Subject: Math            &  0.06     (0.10) & -0.00     (0.08) \\
   Subject: Physics         & -0.07     (0.10) & -0.02     (0.08) \\
   Subject: Other           &  0.08     (0.09) &  0.02     (0.07) \\
   Age                      &  0.00     (0.04) &  0.01     (0.03) \\
   Gender: F                &  0.06     (0.07) & -0.02     (0.06) \\
   Gender: Non-binary       &  0.30     (0.60) &  0.02     (0.61) \\
   Sample: Israel           &  0.09     (0.14) &  0.27$^{*}$   (0.12) \\
   Sample: Japan            & -0.06     (0.15) &  0.20     (0.13) \\
   Sample: Norway           & -0.47$^{***}$ (0.14) &  0.72$^{***}$ (0.11) \\
   Sample: Sweden           & -0.30$^{*}$   (0.14) &  0.50$^{***}$ (0.13) \\
   Sample: USA              & -0.52$^{***}$ (0.15) &  0.77$^{***}$ (0.14) \\
  \midrule
  Num. Observations                &  508             & 508        \\
  Num. Imputations                 & 10               & 10         \\
  R$^2$ (adj.)                     & 20\% (16\%)      & 28\% (24\%)\\     
  \bottomrule
  \multicolumn{3}{l}{\textit{Note:} Statistical significance indicated by $^{***} p < 0.001$, $^{**} p < 0.01$, $^{*} p < 0.05$.}
\end{tabular}
\end{table}

First, we found that both AI-EdTech self-efficacy (i.e., teachers' beliefs in their ability to use and discuss AI-EdTech) and AI understanding (i.e., how well teachers understand AI in lay terms) are associated with the teachers' perceived benefits and concerns about AI-EdTech. Specifically, higher self-efficacy and understanding significantly predicted more perceived benefits and fewer concerns.

Second, we found several teachers' individual cultural values significantly predict their perceived benefits and concerns about AI-EdTech. In particular, higher uncertainty avoidance, stronger long-term orientation, and lower masculinity predicted more perceived benefits of AI-EdTech. In contrast, higher masculinity and higher collectivism predicted more concerns about AI-EdTEch.

Third, we found that more experienced teachers (i.e., years of teaching experience) have fewer concerns about AI-EdTech, but they do not perceive more benefits. Moreover, we did not find evidence that teachers' age, gender identity, level of education, or the subject they teach explain their perceived benefits or concerns about AI-EdTech. The absence of attitudinal variation significantly explained by socio-demographic and subject characteristics is remarkable.

Finally, we found country-level variation across samples in teachers' perceived benefits of and concerns about AI-EdTech. Adjusting for all other variables, teachers in the Brazilian (the reference group in Table~\ref{tab:regression1}), Israeli, and Japanese samples perceived slightly more benefits than those in the Norwegian, Swedish, and American samples. Concerns were higher among teachers surveyed in Israel, Norway, Sweden, and the USA, relative to Brazil and Japan.

\begin{table}[ht]
\centering
\small
\caption{Results for multiple regressions explaining teachers' trust in AI-EdTech using three sets of variables: (1) only perceived benefits and concerns of AI-EdTech, (2) various teacher characteristics and fixed effects for the geographic location (sample), but excluding perceived benefits and concerns, and (3) the combination of variables in models (1) and (2). Regression coefficients are shown with heteroskedasticity-robust standard errors in parentheses.}\label{tab:regression2}
\begin{tabular}{lccc}
  \toprule
   \textbf{Outcome: Trust in AI-EdTech}  &  \textbf{(1)} & \textbf{(2)} & \textbf{(3)} \\
   \midrule
   (Intercept)              &  2.82$^{***}$ (0.19) &  2.06$^{***}$ (0.33) &  2.04$^{***}$ (0.30)  \\
   Benefits of AI-EdTech   &  0.45$^{***}$ (0.03) &                      &  0.39$^{***}$ (0.03) \\ 
   Concerns about AI-EdTech   & -0.36$^{***}$ (0.04) &                      & -0.35$^{***}$ (0.04) \\ 
   AI-EdTech Self-Efficacy            &                      &  0.16$^{***}$ (0.03) &  0.05         (0.03)  \\
   AI Understanding            &                      &  0.10$^{**}$  (0.03) &  0.04         (0.03)  \\
   Uncertainty Avoidance    &                      &  0.14$^{**}$  (0.04) &  0.11$^{**}$  (0.04)  \\
   Masculinity              &                      & -0.02         (0.04) &  0.06$^{*}$   (0.03)  \\
   Collectivism             &                      &  0.03         (0.04) &  0.07$^{*}$   (0.03)  \\
   Long-term Orientation    &                      &  0.12$^{*}$   (0.06) &  0.06         (0.04)  \\
   Power Distance           &                      &  0.05         (0.05) &  0.03         (0.04)  \\
   Experience in Education  &                      & -0.00         (0.04) & -0.04         (0.03)  \\
   Teaching with Technology &                      &  0.02         (0.04) &  0.03         (0.03)  \\
   Highest Edu.: Master     &                      & -0.09         (0.07) & -0.03         (0.05)  \\
   Highest Edu.: Doctorate  &                      & -0.15         (0.12) & -0.17         (0.09)  \\
   Highest Edu.: Other      &                      & -0.05         (0.19) &  0.05         (0.15)  \\
   Subject: Biology         &                      &  0.04         (0.08) & -0.02         (0.07)  \\
   Subject: Chemistry       &                      & -0.03         (0.08) & -0.05         (0.06)  \\
   Subject: CS              &                      & -0.12         (0.08) & -0.07         (0.06)  \\
   Subject: Math            &                      & -0.01         (0.08) & -0.04         (0.06)  \\
   Subject: Physics         &                      & -0.10         (0.08) & -0.08         (0.06)  \\
   Subject: Other           &                      & -0.01         (0.07) & -0.04         (0.06)  \\
   Age                      &                      &  0.01         (0.03) &  0.01         (0.03)  \\
   Gender: F                &                      &  0.06         (0.06) &  0.03         (0.05)  \\
   Gender: Non-binary       &                      &  0.14         (0.60) &  0.03         (0.54)  \\
   Sample: Israel           &                      &  0.18         (0.12) &  0.24$^{**}$  (0.09)  \\
   Sample: Japan            &                      &  0.00         (0.13) &  0.10         (0.10)  \\
   Sample: Norway           &                      & -0.72$^{***}$ (0.12) & -0.28$^{**}$ (0.10)  \\
   Sample: Sweden           &                      & -0.44$^{***}$ (0.12) & -0.15         (0.10)  \\
   Sample: USA              &                      & -0.65$^{***}$ (0.14) & -0.18         (0.11)  \\
  \midrule
  Num. Observations         & 508                  & 508              & 508        \\
  Num. Imputations          & 10                   & 10               & 10         \\
  R$^2$ (adj.)              & 49\% (49\%)          & 30\% (26\%)      & 58\% (56\%) \\     
  \bottomrule
  \multicolumn{4}{l}{\textit{Note:} Statistical significance indicated by $^{***} p < 0.001$, $^{**} p < 0.01$, $^{*} p < 0.05$.}
\end{tabular}
\end{table}

The second set of regression analyses focused on trust in AI-EdTech systematically decomposes the influence of the two proposed antecedents and various other teacher characteristics (Table~\ref{tab:regression2}). Model 1 confirmed that perceived benefits of AI-EdTech and concerns about AI-EdTech are indeed significant predictors of trust in AI-EdTech, and they explain 49\% of variance in trust.

Model 2 examined which teacher characteristics explain trust in AI-EdTech, without consideration of perceived benefits and concerns. It yielded similar results to those predicting the antecedents of trust in Table~\ref{tab:regression1}: both AI-EdTech self-efficacy and AI understanding were significant positive predictors of trust; additionally, higher uncertainty avoidance and a stronger long-term orientation predicted higher trust. Trust was higher in the Brazilian (reference group in Table~\ref{tab:regression2}), Israeli, and Japanese samples, compared to the Norwegian, Swedish, and American samples.

Model 3 examined the combination of predictors from Models 1 and 2 to understand what teacher characteristics matter for trust in AI-EdTech above and beyond perceived benefits and concerns about AI-EdTech. The variance explained by the combined model is 9 percentage points higher (58\%) than the model with only perceived benefits and concerns (49\%). The inclusion of the various teacher characteristics in Model 3 only slightly reduced the magnitude of the significant coefficients on perceived benefits and concerns. In the combined model, AI-EdTech self-efficacy and AI understanding are not significant predictors of trust anymore, suggesting that their variance is already explained by perceived benefits and concerns. Yet we find three significant cultural variables that predict higher trust in AI-EdTech: higher uncertainty avoidance, higher masculinity, and higher collectivism. Finally, in model 3, trust is higher in the Israeli sample and lower in the Norwegian sample, compared to the Brazilian, Japanese, Swedish, and American samples.

\section{Discussion}

We investigated which teacher characteristics shape teachers' trust in AI-EdTech, and its proposed antecedents, perceived benefits of and concerns about AI-EdTech in K-12 education. In summary, our survey research across six geographically and culturally distinctive samples found evidence that teachers' trust in AI-EdTech depends on their self-efficacy with AI-EdTech, their understanding of AI, their cultural values, and their geographic location. We found that all of these factors predicted teachers' perceived benefits and concerns about AI-EdTech, which in turn predicted their trust in it. Overall, this work highlights potential ways to foster teachers' trust in AI-EdTech to facilitate its adoption and effective use in classrooms. Next, we discuss each finding and its implications for theory and practice.

\subsection{Trust Depends on Teachers' Self-efficacy and Understanding}

As depicted in Figure~\ref{fig:schema}, we found that teachers' perceived benefits and concerns about AI-EdTech significantly influence their trust in AI-EdTech in K-12 education. These two antecedents of trust are significantly predicted by teachers' self-efficacy with AI-EdTech and understanding of AI. The regression results suggest that the positive relationship between AI-EdTech self-efficacy and trust is mediated by higher perceived benefits and fewer concerns about AI-EdTech. While prior work has found the importance of self-efficacy for teachers' trust in specific national contexts~\citep[e.g.,][]{cukurova2023adoption}, the current results suggest a mechanism by which AI-EdTech self-efficacy influences teachers' perceptions, which are a basis of their willingness to make themselves vulnerable by using AI-EdTech. The results also show that teachers with a more realistic understanding of AI perceived more benefits and fewer concerns, and ultimately reported more trust. Remarkably, we did not find that teachers' age, gender, level of education, or experience using digital technology in education influenced their trust in AI-EdTech. 
This suggests that educating teachers about AI and especially AI-EdTech can forster more trust at all stages and across all domains. Activities that enhance self-efficacy could be integrated into professional development (PD) or as part of the teacher training curriculum. PD can equip teachers with relevant knowledge about AI-EdTech and address common misconceptions about AI. It does not make teachers AI engineers, but it helps them develop a grounded and realistic understanding of AI as a tool, how it might benefit them, what concerns are warranted, and it informs their willingness to try it out in their classrooms.

Teaching K-12 teachers about AI is a challenging and complex task, and it is not yet prevalent in formal school settings~\citep{casal2023ai}. However, scholars have started exploring ways to implement PD programs, with positive results on teachers' increased understanding of, and trust in, AI-EdTech~\citep[e.g.,][]{nazaretsky2022teachers}. Building on the results of a study by~\cite{celik2023towards}, we suggest that such programs should focus on providing teachers with technological and pedagogical knowledge about AI-EdTech to support effective adoption in K-12 settings and ultimately improve students' learning outcomes. Our results emphasize the importance of emerging AI literacy frameworks for teacher PD~\citep[e.g.,][]{miao2022k}, ethical principles on using AI in K-12 settings~\citep[e.g.,][]{adams2023ethical}, and research on AI literacy~\citep{casal2023ai}. Our results highlight the importance of teachers' understanding of AI and the need to move from unrealistic visions of AI (e.g., autonomous robots as in terminator-like science fiction) to realistic one (e.g., as algorithms to support decision-making based on big data), as this can influence teachers' trust by calibrating their perceived benefits and concerns about AI-EdTech. These results have significant implications on practice and policy-making regarding AI competency development for teachers. Teacher training and PD initiatives should aim to improve basic understanding of AI, rather than provide in-depth courses on AI foundations and techniques. We encourage future research to expand on these findings by examining teacher attitudes and responses to PD with more nuanced measures of teacher understanding, controlling for their prior knowledge and experience in AI as well as in other grade levels, including primary and secondary school settings.

\subsection{Cultural differences in teachers' trust in AI-EdTech}

Culture emerges at many different levels from geographical regions, nations, districts, schools, and classrooms. We found evidence of cultural differences in teachers' trust in AI-EdTech, which suggests that individual cultural values should be considered when integrating AI-EdTech in the K-12 setting. In particular, we found significant differences in trust along four cultural dimensions: uncertainty avoidance, long- vs. short-term orientation, masculinity vs. femininity, and collectivism vs. individualism. This evidence indicates that Hofstede's model of culture~\citep{hofstede2010cultures} is relevant for considerations of the adoption and effective use of AI-EdTech among teachers. Individual cultural values, unlike AI-EdTech self-efficacy and AI understanding, were found to explain variation in trust even when adjusting for perceived benefits and concerns. This suggests that their influence on trust is more fundamental and not sufficiently captured by perceptions of AI-EdTech benefits and concerns.

Our results echo some of the findings of a review of factors that contribute to the acceptance of AI~\citep{kelly2022factors}; the review also finds that culture plays a role in users' acceptance or rejection of AI. However, the findings of the review article are not grounded in studies conducted in K-12 education, and they do not explain exactly which cultural values are critical to consider for the effective adoption of AI-EdTech by teachers. Without this contextualized and specific knowledge, we are limited in our ability to develop culture-sensitive implementations that can facilitate the adoption and use of AI-EdTech by teachers~\citep[e.g.,][]{van2020culture}. Our study findings offer a more nuanced perspective grounded in a sample of teachers across six culturally diverse countries with a wide spectrum of preparedness to harness AI, according to the AI-Readiness Index~\citep{aiready2022}.

First, we found that teachers with a higher level of uncertainty avoidance have more trust in AI-EdTech and perceive it to have more benefits (Tables~\ref{tab:regression1}-\ref{tab:regression2}). The association between uncertainty avoidance and trust is in line with prior work that found the same relationship in the context of automation across the US, Taiwan, and Turkey~\citep{chien2016relation}. In K-12 education, it makes sense that teachers are more likely to adopt AI-EdTech when they experience less uncertainty about different aspects of AI-EdTech. This can be realized by equipping educators with AI-EdTech solutions that are trustworthy, credible, and of high integrity for use in their instructional practices. Furthermore, enhancing this process involves offering detailed guidance and support focused on the transparency and explainability of AI algorithms within AI-EdTech platforms. Such measures can ensure that educators are better informed about the potential and limitations of these tools, including understanding the circumstances under which they might fail. 
This raises questions about the inherently unpredictable characteristics of generative AI technologies. Given the inherent challenges in uncertainty avoidance with these technologies~\citep{bender2021dangers}, their integration into AI-EdTech requires careful investigation. This is especially crucial during the early phases of adoption, where the implications and potential risks must be thoroughly assessed and managed with extra vigilance.

Second, we found that teachers with a stronger long-term orientation trusted AI-EdTech more and perceived more benefits (Tables~\ref{tab:regression1}-\ref{tab:regression2}). Notably, the positive influence of long-term orientation on trust appears to be mediated by perceived benefits (compare models 2 and 3 in Table~\ref{tab:regression2}). 
A long-term orientation is associated with persistence, perseverance, and adaptability~\citep{hofstede2010cultures}, which are all helpful traits for technology adoption. Our findings suggest that it can help to orient teachers to the potential longer-term benefits of integrating AI-EdTech to realize the utility value of additional efforts in the long term. This may be realized through targeted PD, which could highlight the downstream benefits of AI-EdTech in contexts where teachers hold a long-term orientation.

Third, we found that teachers with higher masculinity scores perceived fewer benefits and more concerns regarding AI-EdTech in education. The masculinity dimension can offer insights into how teachers from different cultures perceive and interact with AI in education, potentially impacting their perceived benefits, concerns, and trust in AI-EdTech. The average masculinity scores were below the midpoint (<3 on the Likert scale) for all six country samples in this study (Table~\ref{tab:desc}). However, more masculine societies (e.g., Japan and the USA) might have higher expectations for AI's success and achievement possibilities in education and therefore more concerns if it does not meet these expectations due to the slow adoption of such tools. While masculinity, as well as collectivism, predict concerns about AI-EdTech, these cultural dimensions significantly predict trust only while adjusting for both perceived benefits and concerns, which suggests that their observed influence may not be as robust as that of the other cultural values.

\subsection{Limitations}

This study has several limitations to consider when drawing implications from our findings. First, we collected a convenience sample in each country that is not nationally representative, limiting the generalizability of the result compared to a more costly probability sample~\citep[e.g.,][]{jager2017ii}. To address this concern, we tried our best to reach a heterogeneous teacher sample in each country. 

Second, we emphasize that our results represent a snapshot of teachers' trust in AI-EdTech at a certain point in time, while AI, and its public perception, are highly dynamic. Thus, even without designated interventions such as teacher PD programs, teachers' attitudes are expected to change based on their experiences with various AI-EdTech tools. In addition, such changes can occur in response to significant events that capture widespread attention, such as the sudden popularity of OpenAI's ChatGPT. In this regard, we note that most of our data were collected after the release of ChatGPT, at the initial stages of its use. 

Third, cross-cultural research, like the present study, compares samples that differ not only in their geographic location and culture but also in other, often unobservable characteristics. Further studies can build on this work to test the external validity of our results to the broader population of teachers (other grade levels such as pre-school and primary school) in K-12 education in each country. 

Fourth, while the translation of the survey instrument from English into Swedish, Japanese, Portuguese, Norwegian, and Japanese was performed by native speakers and double- or triple-checked by domain experts, the translated surveys did not undergo formal psychometric evaluation. This is a common limitation in cross-cultural research involving survey instruments~\citep{he2012bias}. 

Finally, we caution against generalizing our results to K-12 contexts without examining their cultural contexts. To this end, we recommend collecting cultural measures that do not rely on Hofstede's original national scores~\cite{hofstede2010cultures}, since cultural values are not static~\cite{varnum2017cultural}, and instead measuring cultural values at an individual level for the target population (i.e., teachers), as we have done in this study.

\subsection{Broader implications and next steps}

The results of this study have implications that transcend beyond simple correlations of various factors influencing teachers' trust in AI, and its two antecedents. Echoing recent calls in the literature~\cite[e.g.,][]{cukurova2023adoption, kizilcec2023advance}, our findings highlight the value of a broader spectrum of considerations at the teacher, classroom, school, and educational ecosystem levels when researching the adoption of AI-EdTech. Specifically, at the teacher level, constructs like AI literacy would benefit from a broader perspective. Although often considered to be limited to technical knowledge acquisition, practical know-how of some AI applications, and straightforward ethical considerations, AI literacy would benefit from educator-centered perspectives that involve considerations of human agency, the increasing need for lifelong learning, as well as a mindset that prioritizes cultural values and human intelligence. We demonstrate that cultural values are significant predictors of teachers' perceived benefits, concerns, and trust in AI-EdTech. Cultural values and their wider implications therefore warrant consideration in the research and practice of AI literacy, teacher training, and PD efforts.

In light of the proliferation of generative AI technologies and the increasing number of teachers using AI tools, we encourage future research to evaluate these aspects based on their actual usage in authentic settings. Such research will be vital for gaining a deeper understanding of the impact of AI-EdTech on teaching and learning and their adoption by educators across countries. Future research could investigate whether teachers' experience with EdTech leads to higher trust in AI, ultimately leading to more effective adoption and use of AI-EdTech in K-12 education. Moreover, future studies might also examine teachers' trust in and attitudes toward using AI, based on the type of AI application, such as decision-making systems or generative AI applications. Moreover, the results of this study demonstrated that certain individual cultural values are influential for teachers' trust in AI-EdTech. To gain a deeper understanding of the possible influence of other cultural values and dimensions, we recommend qualitative research studies and applying other cultural models. Furthermore, for designers of AI-EdTech, we recommend considering the implementation of culture-sensitive design methods that have been used in other contexts~\citep{van2020culture}.

\section{Conclusions}

The results of this study offer a comprehensive, international account of teachers' trust in AI-EdTech and two of its key antecedents. Specifically, our research shows that teachers' perceived benefits and concerns can account for much of their trust in AI-EdTech in K-12 education. We also present novel insights into how teachers' understanding of AI and their AI-EdTech self-efficacy influence their perceived benefits, concerns, and trust in AI-EdTech across six different countries. We further examine geographical and cultural factors in teachers' trust in AI-EdTech and discuss their implications for training and professional development initiatives aimed at enhancing teachers' understanding of and confidence in AI-EdTech. By considering these factors at various levels, including the individual teacher, the school ecosystem, and broader cultural contexts, we can gain a deeper understanding of the adoption of AI-EdTech in schools. This comprehensive approach can pave the way for more informed, ethical, and effective implementation of AI technologies in educational settings. 

\backmatter



\bmhead{Acknowledgments}

We thank our study participants and everyone who has assisted us in translating and distributing the survey. This work was supported by the National Science Foundation (award no. 2237593). 

\bibliography{MAIN}


\end{document}